\documentclass{article}

 \usepackage[nonatbib, preprint]{neurips_2025}

\usepackage{amsmath,amsfonts}
\usepackage{array}
\usepackage{subcaption}
\usepackage{caption}
\usepackage{textcomp}
\usepackage{stfloats}
\usepackage{url}
\usepackage{pdfpages}
\usepackage{verbatim}
\usepackage{graphicx}
\usepackage{chngcntr}
\usepackage{calc}

\usepackage{setspace}

\usepackage{booktabs}
\usepackage{tabularx}
\newcolumntype{Y}{>{\centering\arraybackslash}X}

\usepackage{enumitem}

\usepackage{algorithm}
\usepackage{algpseudocode}

\usepackage{color, colortbl}
\usepackage{colortbl}
\def\rb{\rowcolor{black!5}} 
\def\rw{\rowcolor{white}}
\usepackage{tabulary}

\usepackage[acronym]{glossaries}
\newacronym{sEMG}{sEMG}{surface electromyography}
\newacronym{HD-EMG}{HD-EMG}{high-density electromyography}
\newacronym{EMG}{EMG}{electromyography}
\newacronym{MU}{MU}{motor unit}
\newacronym{MUAP}{MUAP}{motor unit action potential}
\newacronym{PD}{PD}{Proportional Derivative}
\newacronym{ECG}{ECG}{electrocardiogram}
\newacronym{BCI}{BCI}{Brain-Computer Interface}
\newacronym{HMI}{HMI}{Human Machine Interface}
\newacronym{SCI}{SCI}{Spinal Cord Injury}
\newacronym{NN}{NN}{Neural Network}
\newacronym{ML}{ML}{Machine Learning}
\newacronym{DRL}{DRL}{Deep Reinforcement Learning}
\newacronym{TCN}{TCN}{Temporal Convolutional Network}
\newacronym{LSTM}{LSTM}{Long-Short Term Memory}
\newacronym{RMS}{RMS}{Root Mean Square}
\newacronym{PPO}{PPO}{Proximal Policy Optimisation}
\newacronym{RMSE}{RMSE}{Root Mean Squared Error}
\newacronym{FSM}{FSM}{Finite-State Machine}
\newacronym{SVM}{SVM}{Support Vector Machine}
\newacronym{CoM}{CoM}{Centre of Mass}
\newacronym{GPU}{GPU}{graphics processing unit}
\newacronym{std}{$\sigma$}{standard deviation}
\newacronym{IMU}{IMU}{Inertial Measurement Unit}
\newacronym[longplural={degrees of freedom}]{DOF}{DoF}{degree of freedom}
\newacronym{PandO}{P\&O}{Prosthetics and Orthotics}
\newacronym{HRI}{HRI}{Human-Robot Interaction}
\newacronym{AAN}{AAN}{assist-as-needed}
\newacronym{MSk}{MSk}{musculoskeletal}
\newacronym{HDsEMG}{HDsEMG}{high-density surface electromyography}
\newacronym{IZ}{IZ}{innervation zone}
\newacronym{CoV}{CoV}{coefficient of variation}
\newacronym{RF}{RF}{Rectus Femoris}
\newacronym{BF}{BF}{Biceps Femoris}
\newacronym{FE}{FE}{finite element}
\newacronym{TA}{TA}{Tibialis Anterior}
\newacronym{SO}{SO}{Soleus}
\newacronym{RoI}{RoI}{region of interest}
\newacronym{ied}{IED}{inter-electrode distance}
\newacronym{ReLU}{ReLU}{Rectified Linear Unit}
\newacronym{CAD}{CAD}{computer-aided design}
\newacronym{IIT}{IIT}{ Istituto Italiano di Tecnologia}
\newacronym{ZMP}{ZMP}{zero moment point}
\newacronym{MDP}{MDP}{Markov decision process}
\newacronym{HL}{HL}{high-level}
\newacronym{LL}{LL}{low-level}
\newacronym{VR}{VR}{virtual reality}
\newacronym{KDE}{KDE}{kernel density estimation}
\newacronym{PCA}{PCA}{principal component analysis}
\newacronym{EEG}{EEG}{electroencephalography}
\newacronym{CNS}{CNS}{central nervous system}
\newacronym{NBS}{NBS}{Natural BionicS}
\newacronym{RL}{RL}{reinforcement learning}
\newacronym{XLA}{XLA}{accelerated linear algebra}
\newacronym{ASIS}{ASIS}{Anterior Superior Iliac Spine}
\newacronym{SPS}{SPS}{steps per second}

\usepackage[hidelinks]{hyperref}

\usepackage[numbers]{natbib}
\title{Neural Control and Learning of Simulated\\Hand Movements With an EMG-Based\\Closed-Loop Interface}

\author{%
  Balint K.~Hodossy\quad\quad Dario~Farina \\
  Department of Bioengineering\\
  Imperial College London\\
ICL-Meta Wearable Neural Interfaces Research Centre\\
}

\begin{document}

\maketitle
\begin{abstract}
The standard engineering approach when facing uncertainty is modelling. Mixing data from a well-calibrated model with real recordings has led to breakthroughs in many applications of AI, from computer vision to autonomous driving. This type of model-based data augmentation is now beginning to show promising results in biosignal processing as well. However, while these simulated data are necessary, they are not sufficient for virtual neurophysiological experiments. Simply generating neural signals that reproduce a predetermined motor behaviour does not capture the flexibility, variability, and causal structure required to probe neural mechanisms during control tasks.

In this study, we present an in silico neuromechanical model that combines a fully forward musculoskeletal simulation, reinforcement learning, and sequential, online electromyography synthesis. This framework provides not only synchronised kinematics, dynamics, and corresponding neural activity, but also explicitly models feedback and feedforward control in a virtual participant. In this way, online control problems can be represented, as the simulated human adapts its behaviour via a learned RL policy in response to a neural interface. For example, the virtual user can learn hand movements robust to perturbations or the control of a virtual gesture decoder. We illustrate the approach using a gesturing task within a biomechanical hand model, and lay the groundwork for using this technique to evaluate neural controllers, augment training datasets, and generate synthetic data for neurological conditions.

\end{abstract}
\section{Introduction}

The planning and execution of motion form a fundamental part of human intelligence. We explore and influence our environment, communicate with each other and express ourselves with movement. Through neural interfacing, we aim to integrate artificial systems into this process, coordinating and synchronising the behaviour of assistive devices with our own neuromechanics. Yet, each of us has a unique body and an evolving ``language'' of biosignals, which may be further altered by neurological conditions. How can a neural interface learn to interpret these individual patterns and their adaptations when they are disrupted?

One approach is to collect large-scale datasets, built on carefully designed protocols, appropriate instrumentation, and broad, inclusive recruitment, so that AI models can learn to accommodate this inter-individual diversity \cite{kaifosh_generic_2025}. This is a time-consuming process that requires a significant amount of planning and resources. It is further hindered by the need to use \gls{HMI} algorithms and hardware actively under development, particularly when involving closed-loop online control scenarios. This leads to a slow design iteration loop, which must be repeated whenever a new signal modality, control task or user population is considered. For this reason, promising concepts and techniques may fail when deployed in real conditions, if they reach that stage.

Because the development of next generation of neural interfaces requires numerous design iterations and extensive user involvement, simulation has become an increasingly important tool. Simulation of wearable assistive devices, for example, has been validated by transferring to real-world environments, users and devices \cite{han_policy_2022, luo_experiment-free_2024}. This approach could similarly be used to catalyse development of prototype \gls{HMI} hardware and control algorithms. When using simulated neural signals to train and develop \glspl{HMI}, we face an apparent contradiction. On the one hand, a real participant cannot “wear” a virtual \gls{HMI} device, particularly when it is meant to provide mechanical assistance such as an exoskeleton. On the other hand, if we introduce a virtual human model to restore the necessary mechanical interactions, a simple one-way, open-loop generation of biosignals is no longer sufficient: the virtual user must be able to adapt its behaviour to the ongoing dynamics of the simulated body and its interaction with the \gls{HMI}.

To bridge this gap, we present a framework that embeds biosignal synthesis within a closed-loop, online simulation of a \gls{HMI} and its user. As a proof of concept, we model an \gls{EMG}-driven gesture decoding task using a musculoskeletal hand model. The central novelty is that the motor control policy of the virtual user is not fixed, but it continuously reacts to the gesture decoder’s output, adapting the simulated behaviour and thereby altering the generated sEMG signals in order to optimise the \gls{HMI} output. These updated signals are then fed back into the decoder, forming a bidirectional, adaptive closed-loop between the virtual human and the \gls{HMI}. This is in contrast to all current simulation approaches that are based on one-way, open-loop signal generators. In addition to virtual prototyping, our methods are relevant for applications such as data augmentation for data-driven \glspl{HMI}. For example, integrating models of neurological conditions could generate pathological signals, reducing the participation burden on the patient population in preliminary studies.

\subsection{Background}
\subsubsection{Electromyography} 
Surface electromyography (\gls{EMG}) provides an indirect measure of motor commands and is widely used both as a control modality and as a tool to monitor neurological status \cite{merletti_tutorial_2019, biering2006spasticity, tucker_control_2015}. However, \gls{EMG} signals are highly sensitive to the neurophysiological state and recording conditions (e.g., electrode placement, tissue composition, perspiration), leading to substantial inter- and intra-subject variability. This variability hampers the generalisation of pattern recognition and AI-based decoding methods to new users and sessions \cite{asghari_oskoei_myoelectric_2007, xu_cross-user_2024}. Recent work has addressed robustness via domain adaptation, meta-learning, and large multi-participant datasets \cite{zhang_domain_2022, kaifosh_generic_2025}. A complementary strategy is to augment real recordings with synthetic \gls{EMG}. When synthesis spans sufficiently diverse physiological and recording conditions, it can improve controller robustness and even enable transfer from simulation to real-world applications \cite{akkaya_solving_2019}. 

Advances in modelling motor unit action potentials and motor unit discharge patterns now permit efficient generation of plausible \gls{EMG} for prescribed motions \cite{negro_multi-channel_2016, mamidanna_muniverse_2025, ma_conditional_2025, ma_neuromotion_2024, simpetru_myogen_2026}. However, these inverse approaches rely on predetermined behaviour and therefore cannot support predictive, closed-loop virtual experiments with \glspl{HMI}. This motivates the development of forward neuromechanical models that couple neural drive to muscles, muscle dynamics, and physics-based interaction with the environment.

\subsubsection{Musculoskeletal Simulation}
Modelling motor tasks with the human \gls{MSk} system is commonly done by representing body segments as rigid bodies and their joints as a system of articulations. Their configuration is best described in either general or anatomical coordinates, depending on the application \cite{featherstone_rigid_2008, wu_isb_2005, hicks_is_2015}. Physics engines provide a validated way to construct and integrate the mechanics of this simulation state. The open-source gold standard physics engine for biomechanics simulation is Simbody \cite{sherman_simbody_2011}, usually accessed through the OpenSim development environment. However, it faces severe limitations when scaling up to data-heavy requirements, such as those of \gls{RL} \cite{caggiano_myosuite_2022}. MuJoCo is an alternative open-source choice of physics engine, optimised for efficient and scalable execution and versatile and robust contact models relevant for manipulation tasks \cite{todorov_mujoco_2012}. Thanks to these properties, it has been increasingly adopted for biomechanical modelling research with heavy computational costs \cite{fischer_reinforcement_2021, song_deep_2021, wang_myochallenge_2025}, although there are many examples that use options such as Simbody \cite{de_vree_deep_2021} or DART \cite{luo_experiment-free_2024, park_generative_2022, lee_scalable_2019} for similar applications. 

Of particular interest in MuJoCo is its recent implementation of \gls{XLA} operations, allowing the simulation and control logic to be compiled into highly parallelisable \gls{XLA} code \cite{zakka_mjlab_2026}. This allows performing the simulation loop on many instances of the environment simultaneously on the GPU, benefiting data-driven algorithms due to the enhanced rate of experience collection \cite{zakka_mujoco_2025}. In this study, we applied MuJoCo's \gls{XLA} features, and adapted an open-source \gls{MSk} model of the hand from the MyoSim repository in a \gls{RL} environment \cite{wang_myosim_2022}.

A \gls{MSk} model establishes a set of controllable variables. Optimal solutions to a given task may be found through a number of methods ranging from trajectory optimisation, collocation, neuromechanical control models, reinforcement learning, static optimization or computed muscle control \cite{dembia_opensim_2020, de_groote_perspective_2021, wu_adaptive_2017, park_generative_2022, ma_neuromotion_2024}. A large portion of \gls{MSk} control algorithms solve directly for modulating muscle activation \cite{luo_experiment-free_2024, park_generative_2022}. Activation is a quantity that linearly scales the configuration-dependent force output. This attenuates the presence of electromechanical delay in the model, which is a key property of interest in \gls{EMG}-based control \cite{burdet_human_2013}. Instead, it is also possible to control muscle excitation, a signal that scales the rate of change of muscle activation and enforces consistent activation and deactivation dynamics \cite{de_groote_evaluation_2016}. However, this generally increases the complexity and challenge of finding a stable solution.

\subsubsection{Reinforcement Learning}

In this study we used motion-tracking \gls{RL} to determine an optimal control policy for gesturing. Motion-tracking \gls{RL} has several properties that are relevant for this application:
\begin{itemize}
    \item \textbf{Predictive dynamics:} As a forward predictive control policy, it can form closed-loop control and react to perturbations and to changes in the environment, unlike inverse dynamics methods. This is an essential detail for our purposes, as it allows the modelling and evaluation of closed-loop, online control strategies. The presence and form of feedback to the user is known to have significant effect on the behaviour and performance, when compared to the offline evaluation of controllers \cite{scheme_validation_2013}.
    \item \textbf{Reward flexibility:} Defining the reward term for \gls{RL} is much more flexible and robust than constructing a reliable cost function for trajectory optimization \cite{song_reaching_2023}. Furthermore, control can be established from partially observed, noisy sources of information, which can be used to represent the sensory modalities of the user.
    \item \textbf{Low computational cost:} Imitation learning provides a stable, precise and fast learning process compared to other \gls{RL} techniques \cite{peng_amp_2021}. Once learned, robust policies can be evaluated under new or perturbed conditions without having to find a new solution \cite{de_groote_perspective_2021}.
    \item \textbf{Analogue of biological learning:} The exploration and skill acquisition process in \gls{RL} shares strong similarities with the adaptation of motor skills and the ethology of animals and humans, with mechanisms to model adaptation on different timescales \cite{merel_hierarchical_2019, sutton_reinforcement_1998}. The \gls{RL} learning process also inherently incorporates the mitigation of risk and uncertainty in the sensing, control and the environment.
\end{itemize}

A distinguishing property when considering the control of \gls{MSk} systems is the over-actuated and non-linear nature of muscle actuation. Both the delay and the agonist-antagonist configuration hinders exploration of control strategies, which needs to be addressed through some mechanism in the learning \cite{caggiano_myosuite_2022, lee_scalable_2019}. Another aspect requiring attention is the neurophysiological plausibility of the control signals. The neural drive to muscles is described as a synergistic and continuous signal, characteristics that are important to reproduce in simulation. Actuation through muscle excitation, and regularization of activation or excitation are two common methods to encourage plausible spatio-temporal distribution of muscle effort \cite{hicks_is_2015, park_generative_2022}.

\subsection{Contributions}

Our work extends the neuromechanical modelling scheme laid out by \citet{ma_neuromotion_2024} to new behaviours and a new programming framework. In summary, the main contributions of this study are the following:

\begin{itemize}
    
    \item We integrated an online \gls{EMG} synthesis framework in a \gls{RL} environment. \gls{EMG} is synthesised alongside kinematics and dynamics and can be processed by a virtual \gls{HMI}. The \gls{MU} activity is generated from muscle excitation signals instead of muscle activation. The virtual user can react to the output of the HMI that receives the synthesised signals, enabling the modelling of online and adaptive control tasks. Through \gls{RL}, long-term adaptation and skill acquisition is modelled.
    \item Gesture decoders were trained on the generated signals, and were deployed in simulated online prediction tasks. We demonstrate that the virtual participant could then adapt to this decoder in this online, closed-loop prediction setting, achieving benefits such as improved decoding accuracy or reduced muscular effort.
    \item We implemented a virtual gesturing task using an open-source physics engine and \gls{MSk} hand model. We leveraged a massively parallelised, GPU accelerated framework to facilitate experience collection. The environments and learning scripts will be released as a modular open-source package for assisting development in the research on publication of this work. We will include relevant additional functionality not investigated in depth in this manuscript, such as \gls{HDsEMG} synthesis with dynamic, spatial \glspl{MUAP} and noise synthesis.
\end{itemize}

It is important to note that our primary contribution is not the \gls{EMG} decoding system. For decoding, we deliberately employ a simple \gls{TCN} architecture due to its straightforward implementation, using it only to demonstrate that the framework can adapt from the user’s policy to improve decoder performance. The core contribution is the online control learning framework itself, which can be combined with non-\gls{RL} control schemes and a wide range of \gls
{EMG} decoding methods. Similarly, our current goal is not to construct a highly accurate neuromechanical digital twin of a specific individual. Instead, we focus on implementing and benchmarking the underlying mechanisms, which can subsequently be tuned and calibrated to yield more accurate, personalised closed-loop models in future work. 

\section{Methods} 
\begin{figure}
    \centering
    \includegraphics[width=1\linewidth]{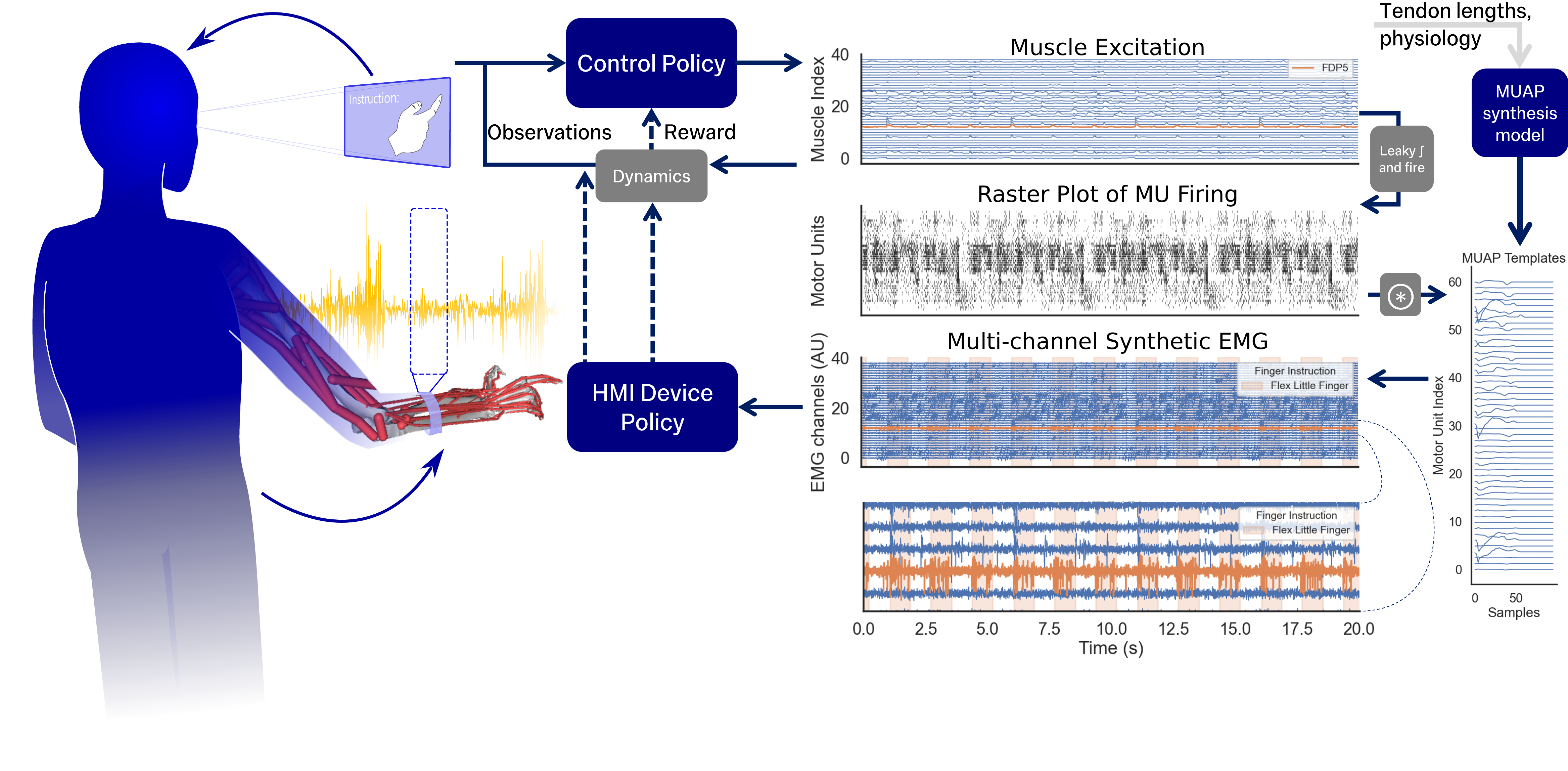}
    \caption{Virtual user in closed-loop HMI control task. Observations are regressed to muscle excitation by the online control policy, which is then associated with a pool of \glspl{MU} with randomised properties. The generated spike trains are convolved with a static set of corresponding \glspl{MUAP} to generate the synthetic \gls{EMG}. We show a 20-s interval, during which the agent cycles through the randomised gesturing pattern 4 times, each finger being flexed and relaxed at different frequencies. For the sake of clarity we visualise only one finger’s flexion and extension patterns indicated by red shaded areas. We highlight the FDP muscle’s corresponding compartment’s activity, the envelope of which is intuitively correlated with the motion.
}
    \label{fig:e-e}
\end{figure}

This section breaks down the implementation of our closed-loop \gls{HMI} simulation environment. First, we describe the MyoHand, the open-source hand model we selected as our starting point, and the modifications we made to adapt it for \gls{EMG} synthesis in parallel computing with JAX \cite{wang_myosim_2022, caggiano_myosuite_2022, bradbury_jax_2018}. Second, we describe our choices for the \gls{RL} used to determine the control policy that defines the behaviour of the virtual user. This is followed by a demonstration of our proof-of-concept closed-loop adaptation model using a simple \gls{TCN} intent decoder. Figure \ref{fig:e-e} illustrates the simulation loops of our framework, including the physics iterations and the closed-loop \gls{EMG} synthesis process via a flowchart.

\subsection{MJX hand environment}

The MyoHand is a complex \gls{MSk} model of the human hand available for the MuJoCo physics engine as part of the MyoSuite project \cite{noauthor_myohubmyosuite_2026}. Its structure and the corresponding parameters were partially adopted from existing OpenSim models, and extended based on physiological studies to better represent advanced manipulation behaviours \cite{caggiano_myosuite_2022}. It consists of:
\begin{itemize}
    \item A kinematic hierarchy including joint transforms, ranges of motion and passive properties. The hierarchy is composed of 29 bones articulated by 23 \glspl{DOF} including the hand, wrist and radioulnar joints. The elbow and every other joint, muscle and segment proximal to it are fixed in place and not simulated.
    \item Inertial properties and collision geometry. Both the fingers and palms produce collisions.
    \item Muscle routing and parameter configurations. 39 muscle-tendon units are used, associated with a 39-dimensional muscle excitation vector which is interpreted as the control space.
    \item Mesh data for visualization of bone configuration. These do not affect simulation and are disabled during training.
\end{itemize}

\begin{figure}
     \captionsetup[subfigure]{
        singlelinecheck = false,
        justification=raggedright
        }
     \centering
    \begin{subfigure}[b]{0.49\textwidth}
        \centering
        \caption{}
        \includegraphics[width=\linewidth]{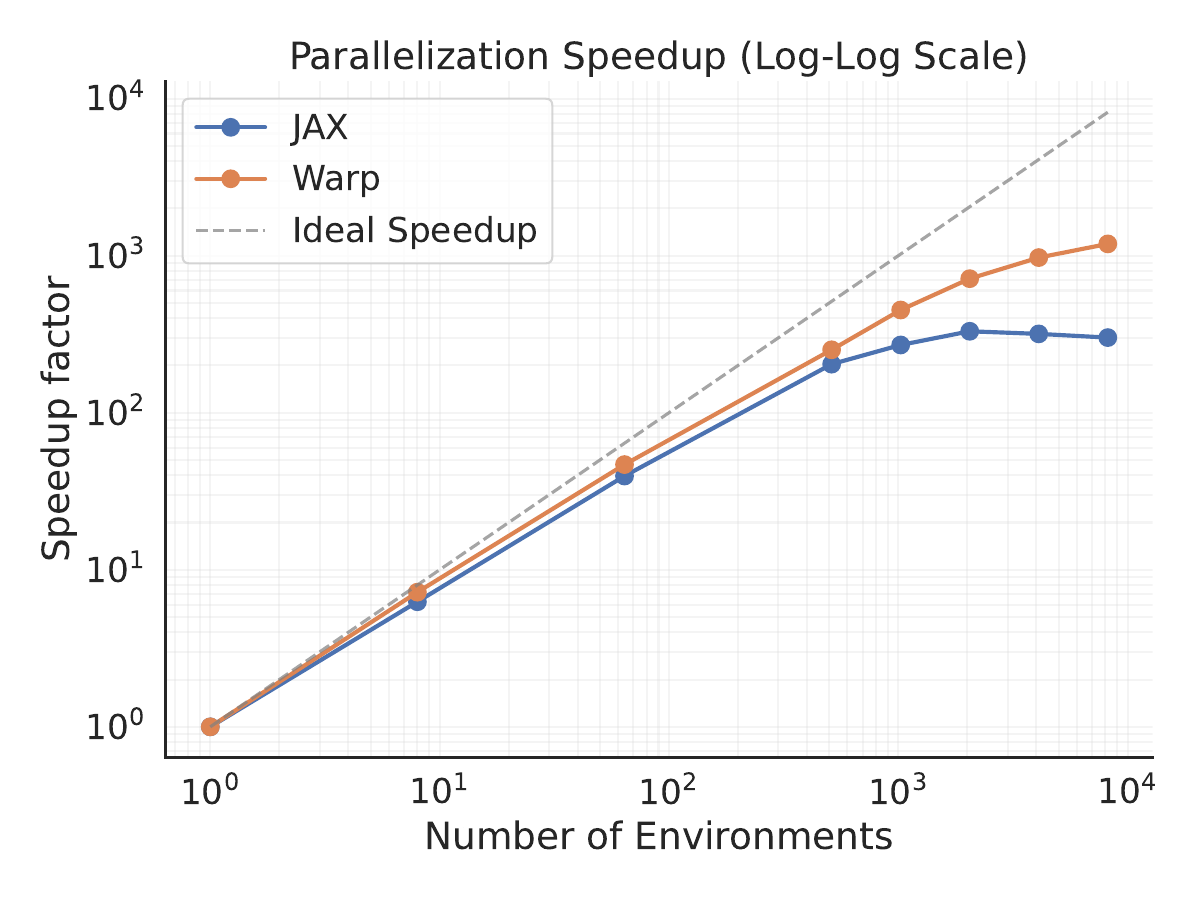}
        \label{fig:sim1}
    \end{subfigure}
     \hfill
     \begin{subfigure}[b]{0.49\textwidth}
        \centering
        \caption{}
        \includegraphics[width=\linewidth]{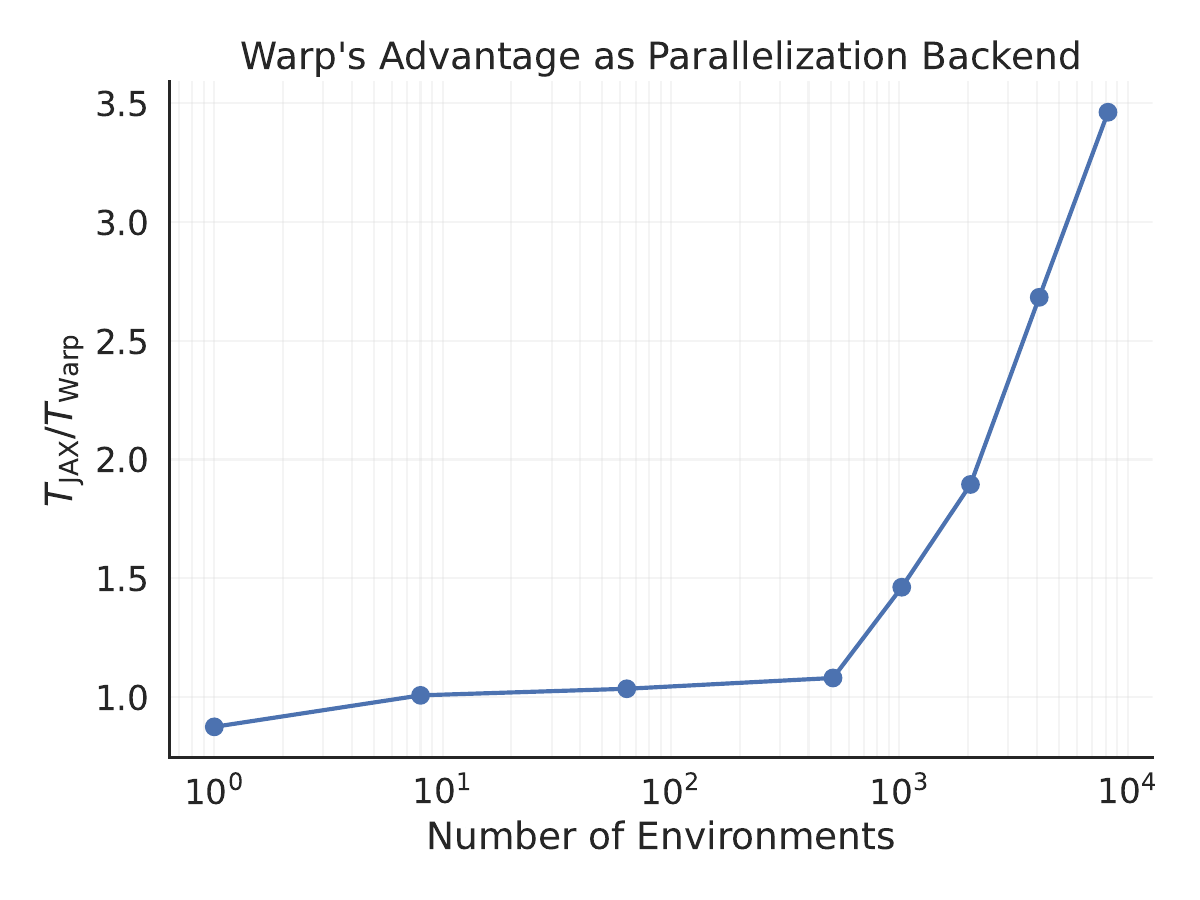}
        \label{fig:sim2}
     \end{subfigure}
    \caption{Performance measures of parallelization with different numbers of environments using two different backends, the generic \gls{XLA} friendly backend (JAX) and the NVIDIA hardware-specific spatial computing toolkit, Warp. These tests were performed with 1024 step long rollouts for all environments on a single NVIDIA L40S \gls{GPU} system with an Intel i7 processor. \emph{a):} Comparing the speedup gained with parallelization using the JAX and Warp frameworks, with respect to the un-parallelised environments built with the corresponding backend. The deviation from the ideal speedup is partially due to Amdahl's law, and partially due to reaching memory and processing bottlenecks. \emph{b):} The ratio of time required to perform the same rollout with JAX or Warp. Particularly for larger environment counts parallelizing the simulation with Warp has increasing benefits.}
    \label{fig:sim}
\end{figure}

To parallelize our simulations, we follow the object-oriented environment scheme presented in MuJoCo Playground \cite{zakka_mujoco_2025}, a collection of parallelized \gls{RL} environments. As a preprocess, we remove elements from the scene that do not affect the simulation to save on memory and compilation time. Following best practices advice for \gls{RL}, we reduce the number of iterations to 4 in Newton constraint solver used, and increase the simulation timestep from 0.002 seconds to 0.004 seconds \cite{noauthor_mujoco_nodate}. A single template object is created from this class. All of this object's features are built using functional programming principles, which allows for the vectorisation of the environment logic to operate on collections of state vectors in a ``single instruction, multiple data'' scheme. This vectorization and compilation to \gls{XLA} operations is done using JAX \cite{bradbury_jax_2018}. Due to memory and processing, as well as Amdahl's law, the benefit from increasing the number of environments does not scale indefinitely. In preliminary testing (see Figure \ref{fig:sim}), we determined 1024 parallel environments to be a suitable compromise. During the development process, another \gls{XLA} compilation option was made available in MuJoCo, based on NVIDIA Warp \cite{noauthor_nvidia_nodate, noauthor_mujoco_nodate}. Warp addresses some limitations of the pure JAX-based backend, primarily when it comes to contacts, leading to significant benefits. In our implementations, we consistently used the more mature JAX backend, but our environments are compatible with Warp and we include some performance comparisons in our results.

Instead of using muscle activation \cite{ma_neuromotion_2024}, we map muscle excitation to the neural drive to stimulate \gls{MU} firing, utilising a parametrised leaky-integrate-and-fire model \cite{teeter_generalized_2018}. These are convolved with a static set of \gls{MUAP} shapes generated using BioMime in the neutral position \cite{ma_conditional_2025}. We provide further implementation details in Appendix \ref{app:EMG}.

\subsection{Reinforcement Learning Policy}

To solve for the controls of the hand model via \gls{RL}, we conceptualise the task as a \gls{MDP}. In this setting, the transition probability to the next environment state is only influenced by the current observed state ($\mathbf{s}_t$), the action sampled by the policy ($\mathbf{a}_t$) and the system dynamics. The performance of the policy is quantified through a scalar reward function $r_t(\mathbf{s_t}, \mathbf{s_{t+1}}, \mathbf{a_t})$. The objective of the optimization is to maximise the expected sum of rewards weighted by a discount factor throughout episodes of learning. Deep learning function estimators with parameters ${\boldsymbol\theta}$ may be utilised to learn the mappings $\pi_{\boldsymbol\theta}(\mathbf{a}_t |\mathbf{s_t})$ describing the action distribution or $v_{\boldsymbol\theta}(\mathbf{s}_t) = \mathbb{E}\{\sum_t\gamma^t r_t\}$ providing a baseline performance estimate \cite{sutton_reinforcement_1998}. We select simple feed-forward neural network estimators for this purpose, the architecture of which will be available in the supplementary materials.

We select \gls{PPO} as our learning algorithm to perform updates to ${\boldsymbol\theta}$, due to its robustness, which makes it a suitable first choice when tackling new tasks \cite{schulman_proximal_2017}. The implementation from \citet{freeman_brax_2021} was used, as it is based on JAX like our environment logic. This allows the two to integrate and be compiled directly together, leading to a significant performance increase \cite{zakka_mujoco_2025}.

\begin{figure}
    \centering
    \includegraphics[width=1\linewidth]{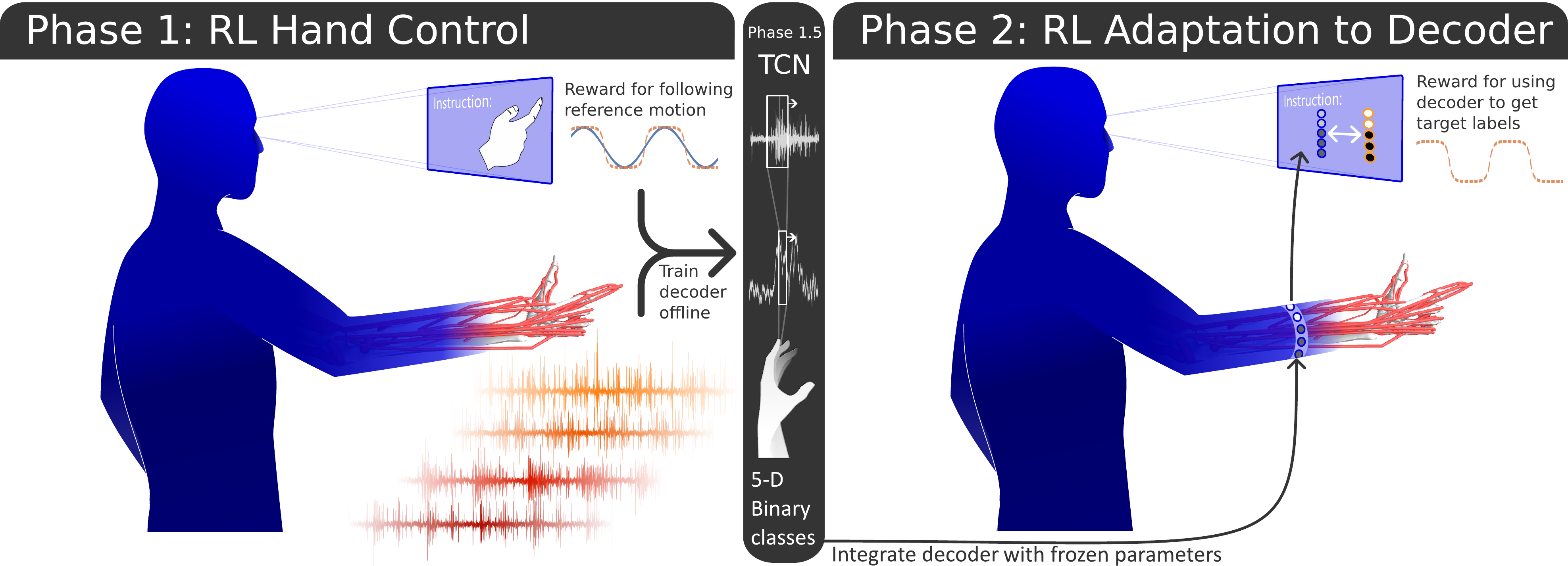}
    \caption{In the first phase of \gls{RL}, a control policy is learnt to follow arbitrary gesture patterns. The synthetic signals from this agent are used to train a virtual subject-specific decoder offline using supervised learning. In the second phase, the agent adapts the motor behaviour to improve the performance with the fixed-parameter decoder, instead of still pursuing an imitation learning goal.}
    \label{fig:explain}
\end{figure}

Before \gls{EMG} synthesis is introduced and the adaptation to a \gls{HMI} can be modelled, we need a baseline virtual user we can develop the closed-loop and adaptation stages with. We select the task of individual finger flexion in arbitrary combinations at various speeds as our goal. The two-phase learning approach we take is illustrated on Figure \ref{fig:explain}.

\subsubsection{First phase: Gesturing task}

\begin{figure}[htbp]
    \centering
    \includegraphics[width=0.65\linewidth]{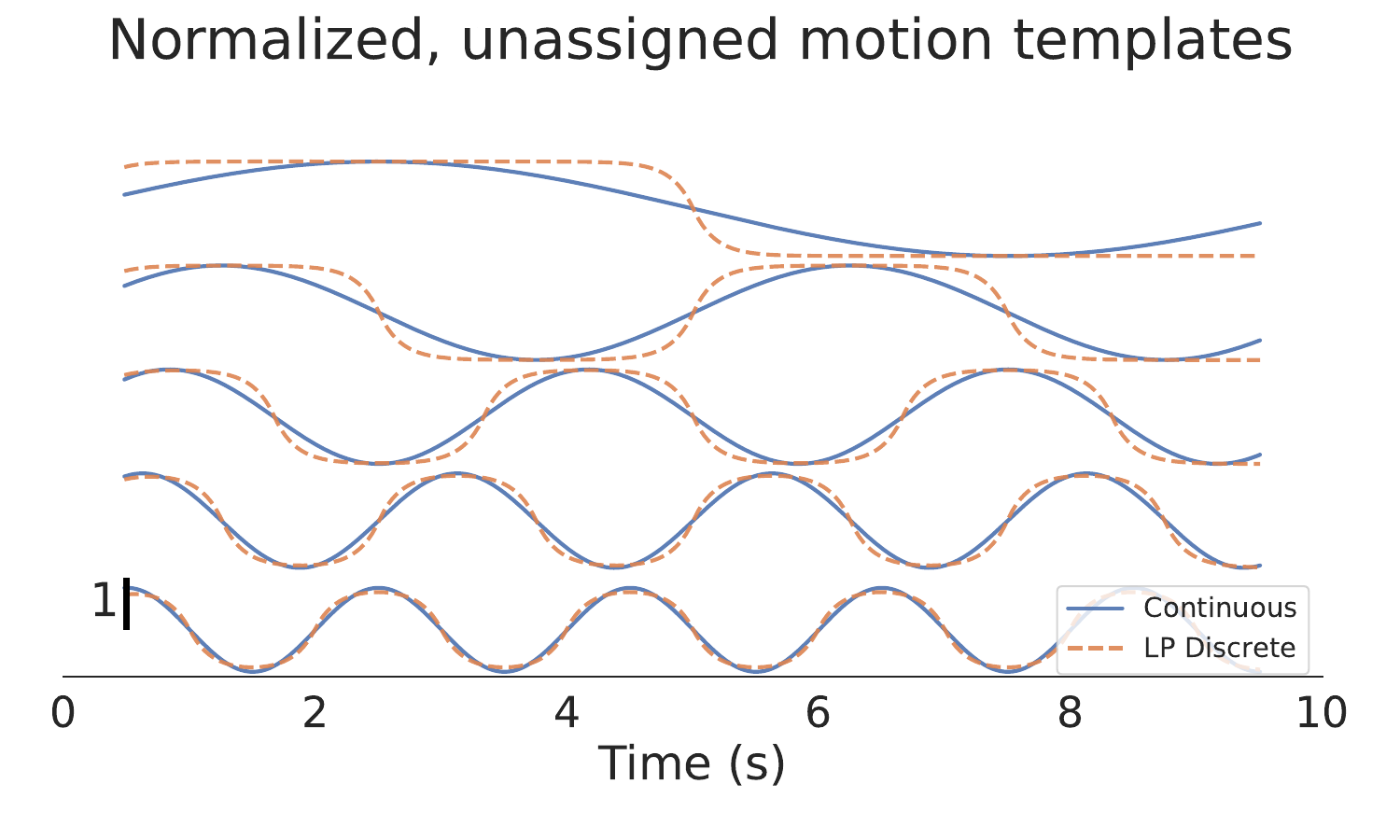}
    \caption{The reference motion for the motion-tracking gesturing task. The templates are shared among all environments, each environment assigns each template to a randomly sampled finger's flexion joints. The template is then scaled and offset as needed according to each respective joint's range of motion. In blue is the purely sinusoidal templates shown, in orange is the discretised then low-pass filtered variant shown. The frequencies are \{0.1, 0.2, 0.3, 0.4, and 0.5\} Hz.}
    \label{fig:motion}
\end{figure}

The parallelization can be leveraged not only to speed up experience collection, but for exploration and regularization of the behaviour. Exploring starts, or reference state initialization is essential for learning robust \gls{RL} policies, and these concepts can be easily integrated into the parallelised learning \cite{peng_deepmimic_2018, sutton_reinforcement_1998}. Instead of using a single motion clip to imitate by our \gls{MSk} control agent, we generate a one template with which a large number of reference behaviours are then produced combinatorially. Sinusoidal motion patterns at five different frequencies are generated (Figure \ref{fig:motion}). These are then assigned to the flexion \glspl{DOF} of each of the five fingers randomly for each environment, as well as randomizing the phase of the motion for each environment. The template curves are scaled and offset to map to 80\% of the joint range of motion. This precalculation means that during simulation getting the reference motion is purely an indexing and scaling operation. Through this, we created a diverse set of motion patterns that a single policy needed to follow for each environment, preventing the memorization and overfitting of a single motion cycle.

During preliminary studies we observed learning to be easier to tune and more robust to perturbation if the motion tracking loss is calculated based on finger endpoint location instead of joint angles directly. Policies converged using either state spaces were compared in terms of recovery times when the pronation \gls{DOF} was perturbed and set to its either joint range maximum. After 100 repetitions each, the fingertip state space policy's recovery to within 15 degrees of the resting pronation angle was $1.14\pm0.27$ seconds, while the joint space policy's time was $1.94\pm0.82$ seconds. For this reason, we performed further experiments with the fingertip-endpoint-based state space. The target endpoints are obtained using forward kinematics on-demand from the reference motion. The motion tracking reward is defined in terms of matching position (Equation \ref{eq:pos}) and velocity (Equation \ref{eq:vel}). 

\begin{equation}\label{eq:pos}
    R_{pos} = w_p \cdot \exp \left( - \sigma_p \cdot \frac{1}{N_{tips}} \sum_{i=1}^{N_{tips}} \| \mathbf{p}_i - \mathbf{\hat{p}}_i \|_2 \right),
\end{equation}

\begin{equation}\label{eq:vel}
    R_{vel} = w_v \cdot \exp \left( - \sigma_v \cdot \frac{1}{N_{tips}} \sum_{i=1}^{N_{tips}} \| \mathbf{v}_i - \mathbf{\hat{v}}_i \|_2 \right),
\end{equation}

where $ \mathbf{p}_i,\mathbf{\hat{p}}_i, \mathbf{v}_i, \mathbf{\hat{v}}_i$ are the reference and current fingertip position and the reference and current fingertip velocities respectively. All other parameters are configurable hyperparameters available in the supplementary materials. The exponential formulation of the reward has the benefit of requiring low error on all terms inside; if the error due to any finger is high, perfectly matching all others will still lead to a low reward, which discourages local optima. We employ a simple cost on the square of the sum excitation to penalise high-effort actions and encourage the distribution of excitation among synergistic muscles (Equation \ref{eq:cost}):

\begin{equation}\label{eq:cost}
    C_{ctrl} = w_c \cdot \frac{1}{N_{ctrl}} \sum_{j=1}^{N_{ctrl}} (\sigma_c \cdot a_j)^2,
\end{equation}
where $a_j$ is the excitation for the $j$-th muscletendon unit determined by the policy. These are then combined to the total reward (Equation \ref{eq:rew})

\begin{equation}\label{eq:rew}
    R_{tot} = R_{pos} + R_{vel} - C_{ctrl}
\end{equation}

\begin{figure}[htbp]
    \centering
    \includegraphics[width=0.5\linewidth]{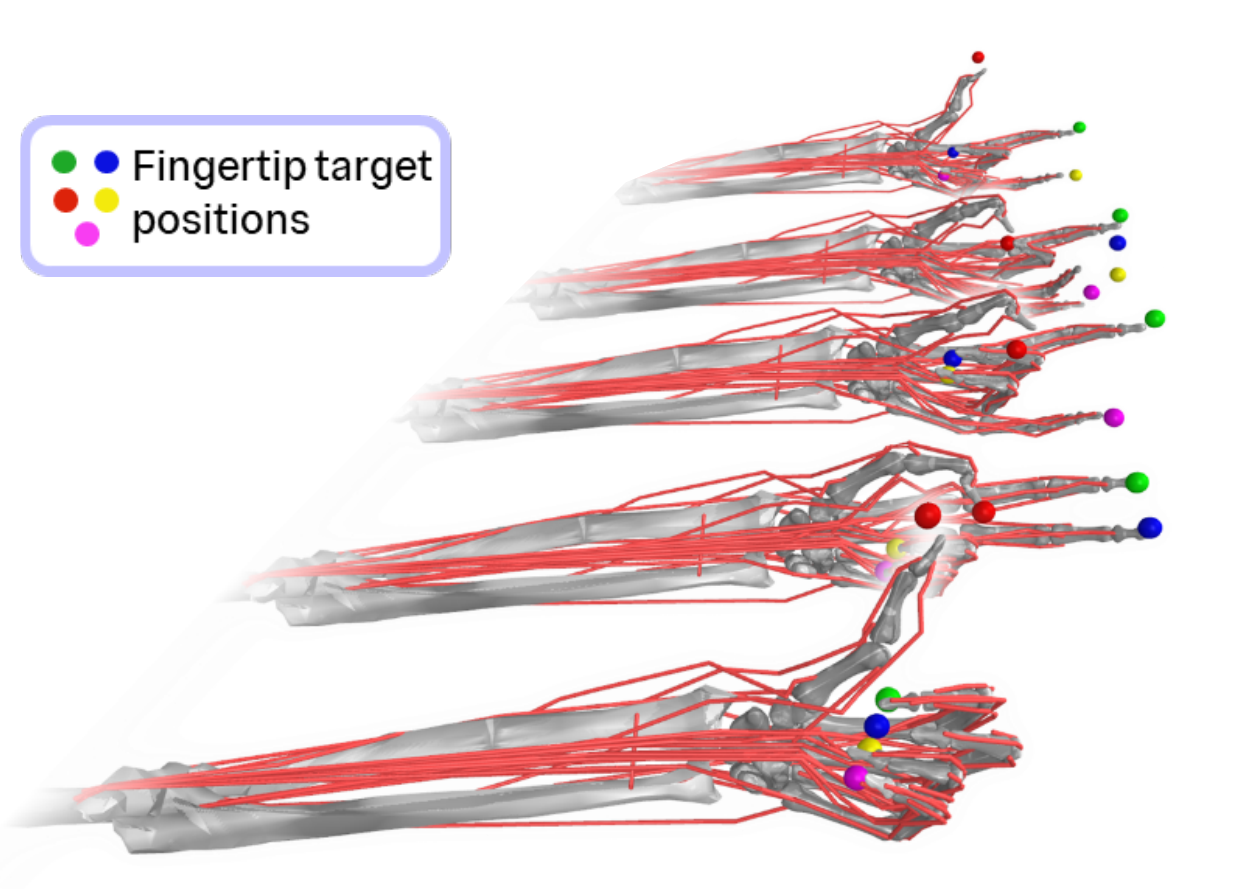}
    \caption{Illustration of the experience collection with parallel environments. Each environment is using a different projection of the reference motion templates shown in Figure \ref{fig:motion}, with different motion frequencies assigned to each finger.}
    \label{fig:par}
\end{figure}

Figure \ref{fig:par} illustrates the motion tracking process.

As gesturing is usually performed in a more discrete manner (e.g., sign language), we also provide the option to use a square-wave signal as template. To aid the extraction of velocity measures and to make the patterns continuous, a low-pass filter or a smoothing function can be applied to the square waves. We used a 1 Hz cutoff frequency and used a zero-phase 2\textsuperscript{nd} order Butterworth filter for this purpose. Figure \ref{fig:motion} presents the resulting motion curves. In our experiments we primarily trained a policy to track the continuous motion and then changed the reference motion to the more challenging discrete movements in a curriculum-like fashion. We provide ablation studies showing similar results are achievable if the discrete behaviour is learnt directly.

The observation vector $\mathbf{o}_t$ is provided to the policy at each step, containing proprioceptive signals, as well as information about the target motion (Equation \ref{eq:obs}).
\begin{equation} \label{eq:obs}
    \mathbf{s}_t = \begin{bmatrix} 
        \mathbf{q} \\ 
        \mathbf{\dot{q}} \\ 
        \mathbf{P}_{ref} \\ 
        \mathbf{P}_{cur} \\ 
        \mathbf{P}_{err} \\ 
        \mathbf{\boldsymbol{\alpha}} 
    \end{bmatrix},
\end{equation}
where $\mathbf{q}, \mathbf{\dot{q}}$ are the joint angles and velocities, $\mathbf{P}_{ref}, \mathbf{P}_{cur}$ are the reference and current Cartesian fingertip positions respectively. We include the spatial error vector as well $\mathbf{P}_{err}=\mathbf{P}_{ref} - \mathbf{P}_{cur}$, as it has been shown to speed up learning in similar tasks \cite{bergamin_drecon_2019}. Lastly, $\boldsymbol{\alpha}$ corresponds to the integrated muscle activations that are scaling the muscle force output.

We include an early termination condition if the angle tracking error average across \glspl{DOF} exceeds hyperparameter $\epsilon_{term}$. Early termination is an essential component for motion tracking tasks to \mbox{catalyse} learning by focusing recovery strategies to near optimal states \cite{peng_deepmimic_2018}. 
\begin{equation}
    d_t = \begin{cases} 
        1 & \text{if } \frac{1}{n_v} \sum_{j} |q_j - \hat{q}_j| > \epsilon_{term} \\ 
        0 & \text{otherwise} 
    \end{cases}
\end{equation}
If this condition is reached the state is deemed terminal and the environment is reset. Otherwise, if an episode exceeds a maximum length (chosen to be 3000 steps in our case), the state value is bootstrapped and is also reset. On reset, we initialise the environment state to a random phase of the reference motion, and apply a small joint angle perturbation (0.06 radians for each \gls{DOF}).

To speed up the learning process, we subsample policy updates to once every 4 simulation steps, between which the muscle excitation levels are kept at the same value provided by the policy the last time it was queried. As we transition the model to the second phase, we reduce this subsampling to once every 2 simulation steps, to have higher temporal resolution of the \gls{MU} activity updates.

\subsubsection{Second phase: EMG-based control task}

As an example \gls{EMG} decoding model, we selected a \gls{TCN}, a simple architecture that has been shown in the past to be applicable to a diverse set of \gls{EMG} decoding tasks \cite{hodossy_high-level_2025, betthauser_stable_2020}. The architecture and hyperparameters are available in the supplementary materials. The original policy is stochastic, but performance improvements can be reached during inference with motion tracking policies by converting them to a deterministic format; this can be achieved by simply using the action distribution's mode. We performed a 50 second rollout with the deterministic policy in 120 environments to generate a training dataset, and reserved 5\% of the data for testing purposes. We then fit the model to a classification task, labelling each element of a 5-dimensional binary vector corresponding to the discretised flexion state of the reference /intended motion recorded concurrently with the synthetic \gls{EMG}. We used the Adam optimiser available from the Optax package, and the neural network operators from the Flax package \cite{deepmind2020jax}. Beyond neural noise, electrode noise may be added to the \gls{EMG} signals to increase its realism and the difficulty of the decoding challenge. Creating sophisticated models of noise and interference alongside techniques to tackle them is a promising application of simulated environments. However, in this study we did not pursue this line of inquiry beyond verifying the decoder learning is robust against white Gaussian noise up until its \gls{std} is comparable with the signal, after which its performance diminishes as expected. 

In the second phase of \gls{RL}, we took this offline-trained \gls{EMG} decoder and embedded it into the learning environment, which was once again governed by a stochastic policy. Now, at each time step, in addition to applying the actions from the \gls{MSk} control policy, we updated a 250 ms long rolling window of the synthesised \gls{EMG} and queried the \gls{EMG} decoder model with it (see Figure \ref{fig:e-e}). In this new task, we no longer require the \gls{MSk} control agent to follow along the reference motion. To encourage improvement of the agent's use of the \gls{EMG} decoder, we alter the reward function, and replace $R_{pos}$ and $R_{vel}$ of the reward function (Equation \ref{eq:rew}) with the expression shown in Equation \ref{eq:new_rew}

\begin{equation}\label{eq:new_rew}
    R_{acc} = w_{acc} \left( 1 - \frac{2}{N_{finger}} \sum_{i=1}^{N_{finger}} |y_i - \hat{y}_i| \right),
\end{equation}

where $y_i,\hat{y}_i$ are the ground truth and predicted labels of the reference motion's flexion state respectively. The ground truth labels are extracted by thresholding the corresponding finger's reference motion template depending on which joint range extrema it is currently closer to.

In a separate condition (see Figure \ref{fig:obs-acc}), we also augment the observation vector by concatenating the vector $\hat{y}_i$. By providing the agent with information about when and for which finger the \gls{EMG} decoder is inaccurate, we reinforce the closed-loop nature of the control, which we hypothesize will enhance the adaptation process.

\section{Results}\label{sec:res}

\subsection{Simulation performance}
We report performance benchmarks describing the benefits of hardware-accelerated and parallelised simulation. The results reported are with a single NVIDIA L40S GPU and an Intel Xeon CPU, and using 1024 parallel environments unless otherwise specified. The exact results reported here are highly dependent on the available hardware and computational load, however, the performance ratios between conditions should indicate possible speed-ups on similar systems.

Average simulation rate with the JAX backend, including environment logic, \gls{MSk} control policy, \gls{EMG} synthesis, \gls{EMG} decoding and \gls{RL} overheads was 110,000 steps per second. Based on the simulation step-size (4 ms) to real-life, the process has a 440-fold speedup factor compared to real life. It is worth noting that for a single environment, a CPU-oriented implementation would greatly outperform the unparallelised GPU-compiled simulation. However, for applications that can take advantage of aggregated experience such as \gls{RL}, the parallelised method is extremely beneficial. 

These are significant improvements over alternatives; for example, in Table \ref{tab:learning-sps} we compare with an existing learning environment for the MyoHand available in the MyoSuite package (without any \gls{EMG} synthesis or decoding). Neuromotion, a similar but open-loop neuromechanical \gls{EMG} synthesis technique, relies on static optimisation with OpenSim \cite{ma_neuromotion_2024}. For hand models, it is possible to find forward mechanics solutions faster than real time with SimBody \cite{blana_real-time_2017, liu_real-time_2022}, however, this comparison does not include control, environment logic or learning. For our approach, the switch to MuJoCo was necessary to make experience collection feasible.

\begingroup
\setlength\extrarowheight{-2pt}
\def\arraystretch{2}
\begin{table}[htbp]
\centering
\caption{Simulation \gls{SPS} across \gls{XLA} and CPU computing backends and environments including learning overheads.}
\begin{tabularx}{1\linewidth}{XXXX|c}\toprule 
Physics Backend & Environment Backend & Learning \mbox{framework} & \mbox{Number of} \mbox{environments} & SPS \\ \midrule \rb
JAX (XLA)& JAX  & Brax & 1024 &  110,000 \\ \rw
CPU (C++)& Python & SB3 + PyTorch & 12 & 1,000 \\ 
\bottomrule
\end{tabularx}
\label{tab:learning-sps}
\end{table}
\endgroup
\subsection{Learning Gesturing}
\begin{figure}[H]
     \captionsetup[subfigure]{
        singlelinecheck = false,
        justification=raggedright
        }
     \centering
     \begin{subfigure}[b]{0.49\textwidth}
        \centering
        \caption{}
        \includegraphics[width=\linewidth]{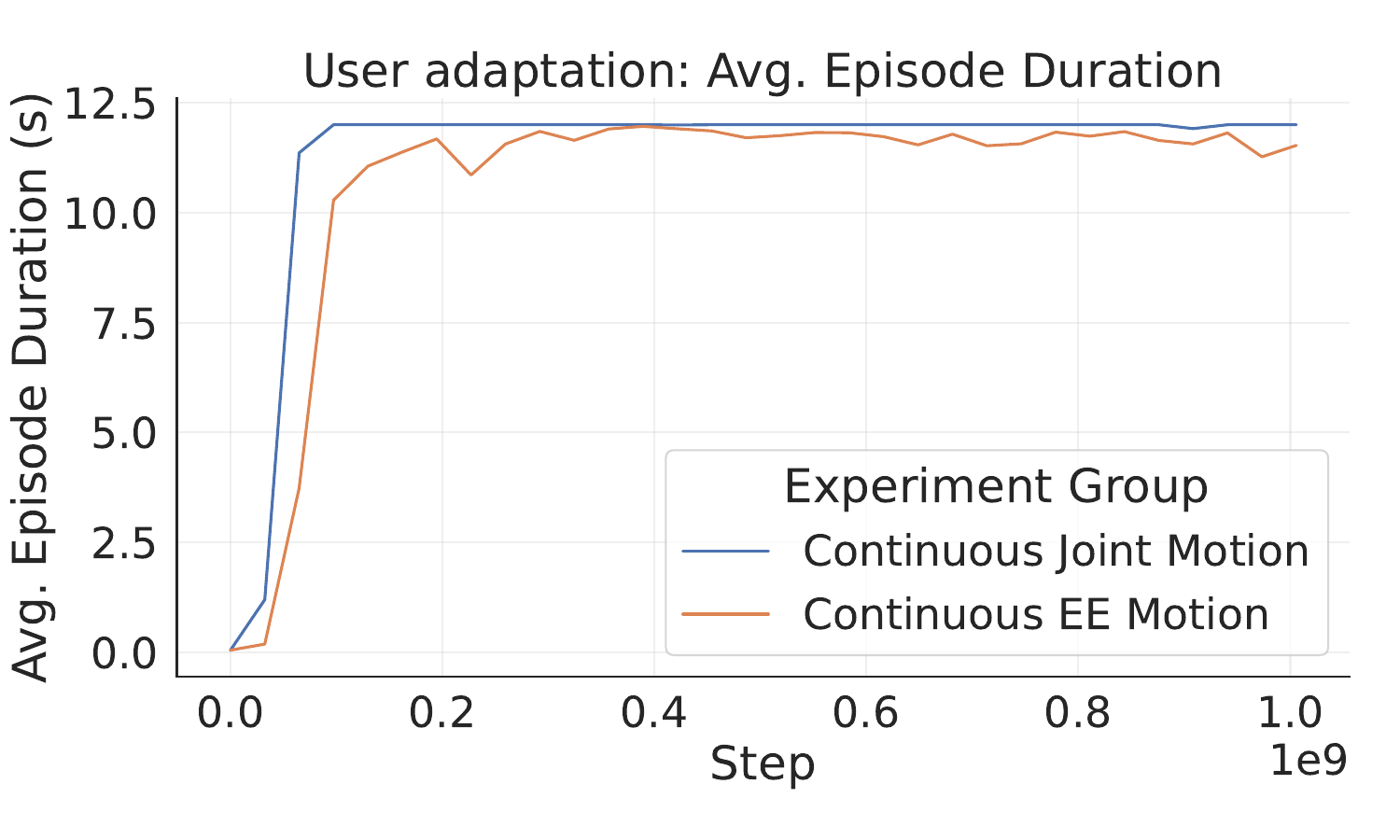}
        \label{fig:rew-dur}
     \end{subfigure}
     \hfill
     \begin{subfigure}[b]{0.49\textwidth}
        \centering
        \caption{}
        \includegraphics[width=\linewidth]{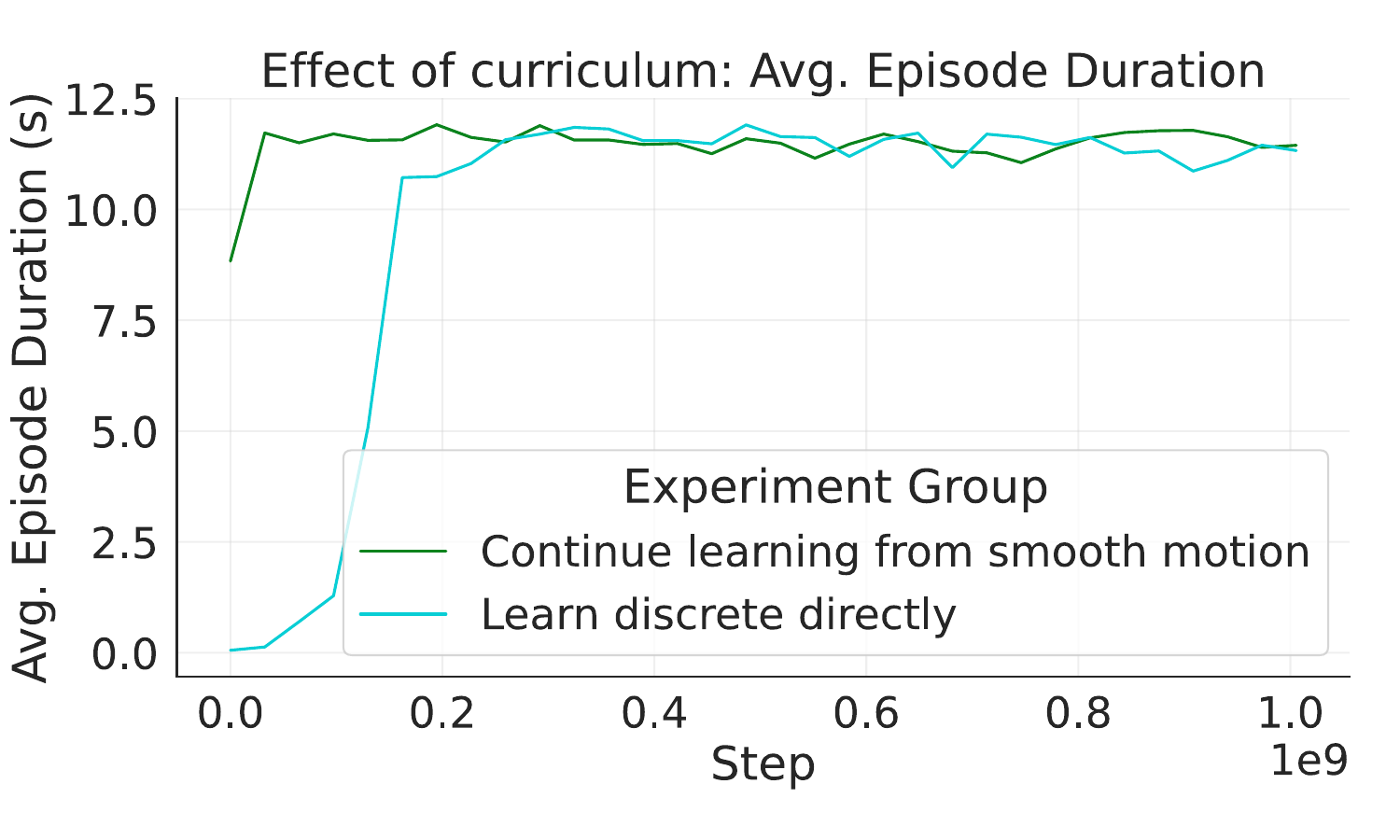}
        \label{fig:disc-dur}
     \end{subfigure}
     
        \caption{Ablation studies on the choice of reward space (\emph{a}) and curriculum learning for the gesture tracking (\emph{b}). \emph{a)} Average episode lengths during learning with the end-effector-based reward and observation spaces (orange) vs. the traditional joint-space approach (blue). The early termination conditions were identical and based on a threshold on joint angle error norm. \emph{b)} Transferring a policy from the continuous gesture task to the discrete one (green) vs. a policy trained directly on the discrete task, also visualised through average episode duration with early termination.}
        \label{fig:ablation}
\end{figure}

Figure \ref{fig:ablation} presents learning curves of the first phase of learning, showing comparisons with alternative design choices of joint-space state representation and non-curriculum learning of motion tracking. Despite the joint-space policy's better alignment with the early termination condition, which leads to better performance in terms of episode length, the policy with the end-effector space is more robust against perturbations.

\subsection{EMG decoder and adaptation}\label{sec:conditions}

\begin{figure}
\captionsetup[subfigure]{
        singlelinecheck = false,
        justification=raggedright
        }
     \centering
     \begin{subfigure}[b]{0.49\textwidth}
        \centering
        \caption{}
        \includegraphics[width=\linewidth]{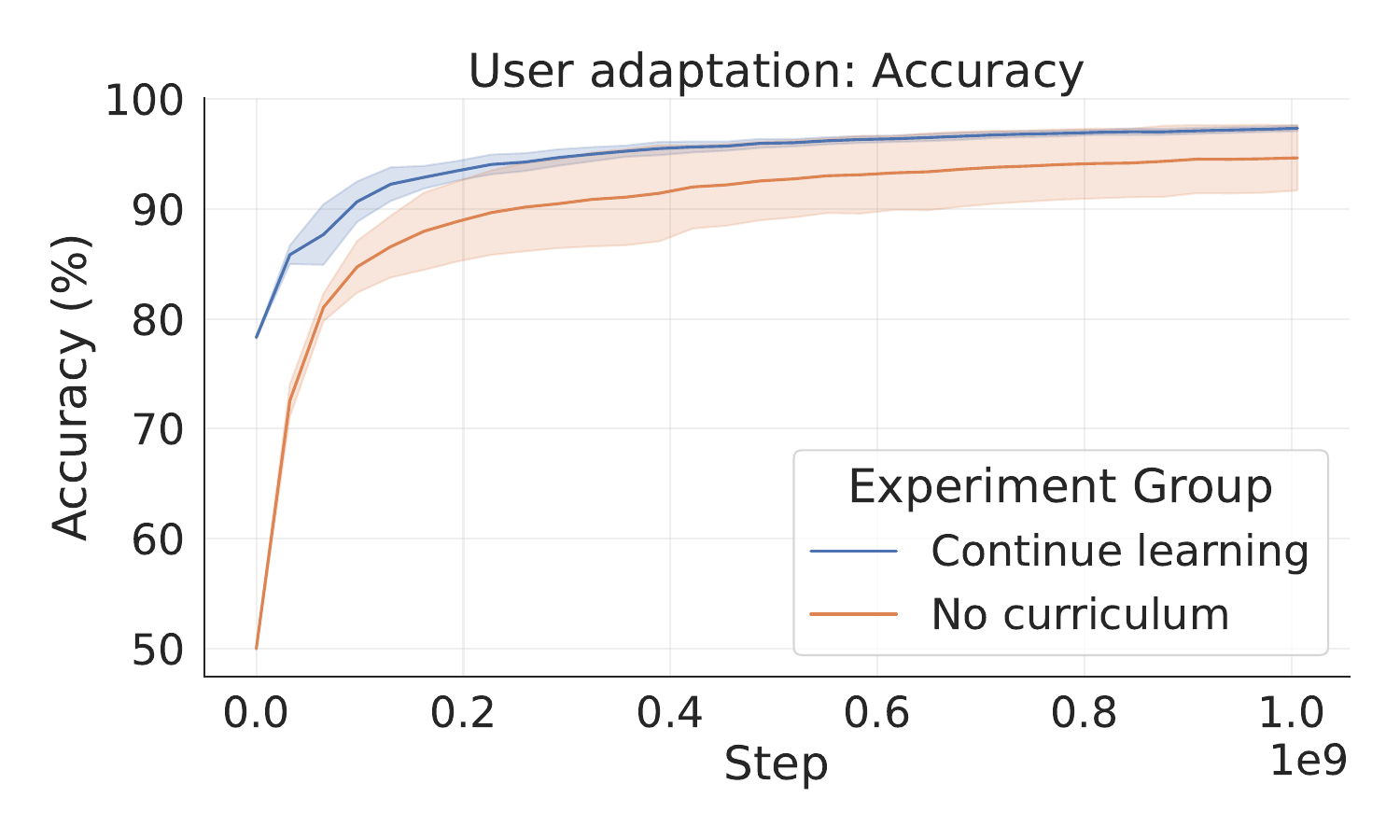}
        \label{fig:curr-acc}
     \end{subfigure}
     \hfill
     \begin{subfigure}[b]{0.49\textwidth}
        \centering
        \caption{}
        \includegraphics[width=\linewidth]{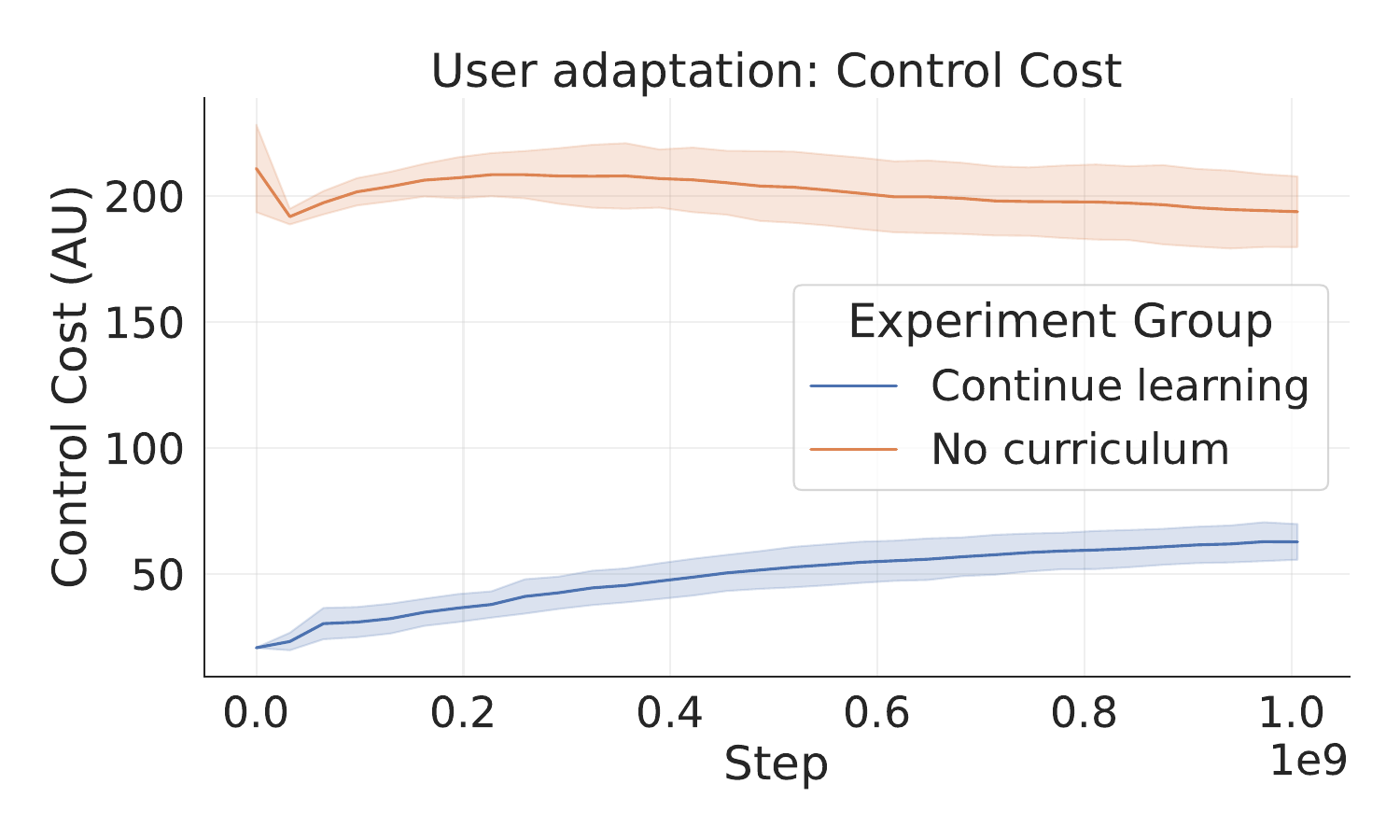}
        \label{fig:curr-cost}
     \end{subfigure}

     \begin{subfigure}[b]{0.49\textwidth}
        \centering
        \caption{}
        \includegraphics[width=\linewidth]{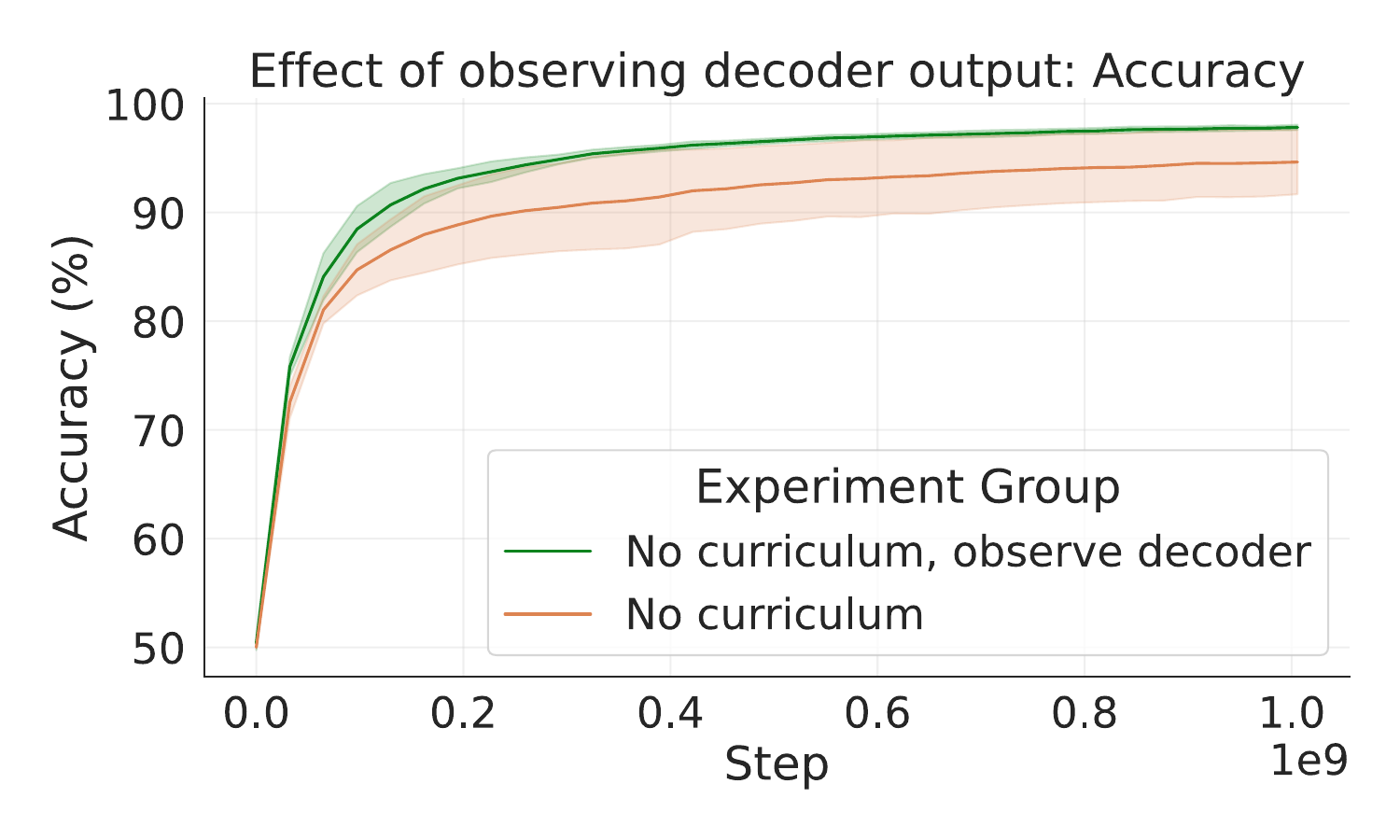}
        \label{fig:obs-acc}
     \end{subfigure}
     \hfill
     \begin{subfigure}[b]{0.49\textwidth}
        \centering
        \caption{}
        \includegraphics[width=\linewidth]{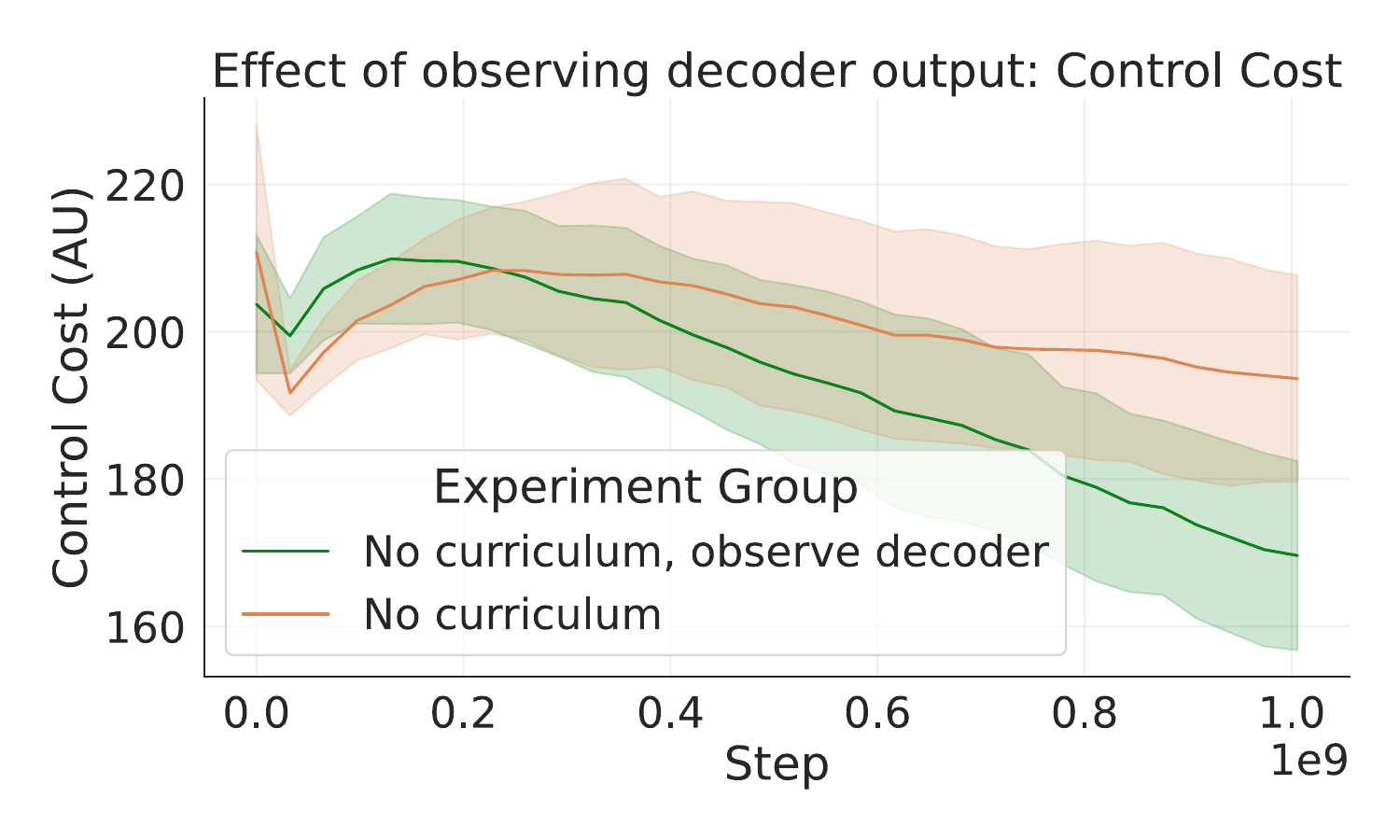}
        \label{fig:obs-cost}
     \end{subfigure}
     
        \caption{Evolution of the reward terms for the \gls{EMG}-decoder-in-the-loop model of user adaptation in the three different conditions described in Section \ref{sec:conditions}. Learning experiments repeated 5 times for all conditions with different randomisation seeds, $\pm$\gls{std} shown in the shaded area. \emph{a)} Accuracy of the pretrained decoder as the virtual user learns to adapt behaviour to improve over time. Average accuracy across the five fingers with two classes (flexed and extended) shown. Condition 1 (blue), overlaid with condition 2 (orange). \emph{b)} Change of the scaled absolute muscle effort over time for the same two policies. \emph{c)} Condition 3 (green) compared with condition 2 (orange) \emph{d)} Effect of the extended observation space on muscle effort over the course of learning.}
        \label{fig:user}
\end{figure}

The \gls{EMG} decoder \gls{TCN}'s training converged in six epochs to an average per-finger classification accuracy of 91\%. To better characterise the virtual user's closed-loop adaptation and to examine the impact randomisation has on the outcome, we repeated the second phase learning five times. In Figure \ref{fig:user} we show the resulting learning curves of three conditions:
\begin{enumerate}
    \item Taking the stochastic policy from phase 1 and continuing learning with the new reward function of Equation \ref{eq:new_rew}.
    \item Reinitialising a new policy independently of phase 1, and learning with the new reward function.
    \item Reinitializing a new policy and in addition to the reward function, also augmenting the observation vector to include the decoder output.
\end{enumerate}
These conditions have different rates of learning and steady state performance. A summary is shown in Table \ref{tab:convergence}.

\begingroup
\setlength\extrarowheight{-2pt}
\def\arraystretch{2}
\begin{table}[htbp]
\centering
\caption{Performance comparison and training steps of the average learning curves of the three conditions described in Section \ref{sec:conditions}.}
\begin{tabularx}{1\linewidth}{X|X|c} \toprule 
Condition & 95\% of max reached (Million Steps) & Max performance \\ \midrule \rb
1. Continue learning & 162.2 & 97.33\% \\ \rw
2. No curriculum & 259.5 & 94.64\% \\ \rb
3. No curriculum, observe decoder & 194.6 & 97.82\% \\ 
\bottomrule
\end{tabularx}
\label{tab:convergence}
\end{table}
\endgroup

\section{Discussion}

In this study, we demonstrated a closed-loop neuromechanical simulation, where the synthetic \gls{EMG} surrogate signal directly influences the behaviour, and vice versa. We designed a gesturing task with a \gls{MSk} hand embedded in a physics engine to take advantage of parallel computing frameworks for efficient experience collection. We used \gls{RL} to learn a control policy for a virtual user, which was used to generate concurrent \gls{EMG} signals through a leaky-integrate-and-fire \gls{MU} model and static \gls{MUAP} templates. The concurrency of the \gls{EMG} synthesis is a crucial requirement to model online control and adaptation, which we exemplified by introducing a pretrained \gls{EMG} decoder to the learning environment. Our results validate that embedding \gls{EMG} synthesis within a reactive, closed-loop simulation allows for the emergence of co-adaptive behaviours that an open-loop approach cannot capture. In the section below we reflect on the result and implementation details, discuss the limitations of the current framework and outline future improvements and applications.

\subsection{Learning and adaptation}
Depending on the conditions, we see different strategies in the adaptation to the \gls{HMI}-use task in the second phase of \gls{RL}. When we continue from the first phase we see the muscle effort (and therefore \gls{EMG} power) increases from its previously converged value over the course of the training, a cost overcome by the gains in the reward term promoting the accuracy with the \gls{EMG} decoder. On the other hand, if the policy is learned from scratch, we see a significantly higher overall muscle activation, but one that is reduced over time. The steady-state value, the rate of learning and the trade-off between decoding accuracy and muscle effort can be tuned by configuring the weight parameters in Equations \ref{eq:new_rew} and \ref{eq:cost}.

It is worth noting that both training on the original gesturing task and closed-loop aid in converging to the optimal solution. In condition 2 (as described in Section \ref{sec:conditions}) the performance is much more dependent on the success of early exploration, as it both starts farther from the solution and it lacks the observation providing the feedback to guide the behaviour. We observe that the adapted policies, particularly conditions 2 and 3, differ considerably in the movement of the wrist \glspl{DOF} from the phase 1 policy, which is an expected side-effect of removing the motion tracking reward. Despite this, all adapted conditions exhibit finger flexion-extension in sync with the target.

Due to our use of \gls{RL} to find optimal behaviours and the ability to generate high signal-to-noise-ratio signals, we may also consider these methods as means to determine practical estimates of the upper bound of performance for different electrode setups and decoding algorithms, as determined by the information transfer of the \gls{HMI}. As such, one application is to use the metrics of the simulated \gls{HMI} to contextualise the performance of real systems, or to compare different systems in simulation.

\subsection{Parallelised learning}
The accelerated experience collection enabled an iterative process of development, which would not have been practical with the traditional frameworks due to the amount of time before the results of a virtual experiment can be evaluated. This comes at a cost of needing to conform to the design requirements of \gls{XLA} and functional programming. These limitations complicate the initial phase of development, as they involve strict management of the flow of environment logic and state with compilation overheads. There are ongoing efforts in the community to alleviate this burden through standardised interfaces, which are necessary to foolproof the process of creating new learning tasks and to reduce the risk of programming errors \cite{zakka_mjlab_2026, zakka_mujoco_2025}.

Despite using a high-performance GPU to generate our final results on an L40S, much of our preliminary experimentation was performed with a consumer-grade RTX 4060 Ti GPU. All of our tooling is compatible with these more accessible devices, albeit at approximately 20\% of the simulation rate when including physics and \gls{RL} overheads; replicating and extending our results is feasible with moderate computational resources, particularly if using the NVIDIA Warp implementation of the physics (see Figure \ref{fig:sim}).

\subsection{Open-source framework}
Our documented and modularised codebase will be released open-source following the publication of this work. Our goal is to collaborate with the community to implement improvements and refinements to our methods, and to make our methods of parallelisable neuromechanical simulation extendable for other applications (e.g. orthotics control or \gls{EMG} synthesis during locomotion).

\subsection{Limitations and future work}
The presented framework should be considered a proof-of-concept, made possible by newly available tools and hardware. Many assumptions with limitations were made in our implementations. Wherever possible, we made these configurable or extensible for future improvements.

\subsubsection{Limitations in the EMG synthesis}
Compared to previous methods, we used muscle excitation as the excitatory signal for \gls{MU} instead of muscle activation \cite{ma_neuromotion_2024}. This is an improvement, as it allows enforcing inherent constraints on the muscle activation dynamics, and because it simulates the electromechanical delay, which would be absent otherwise. However, further improvements are possible to improve the accuracy of the synthesis.

\gls{MU} excitation and inhibition could be modelled as a separate stage, instead of assuming a linear and concurrent relationship with muscle excitation. The control policy could modulate synaptic input to pools of \glspl{MU} explicitly, playing the role of motor cortex and cerebellum \cite{del_vecchio_human_2019}. The resulting discharges could then produce the aggregate muscle excitation, which in turn stimulates muscle activation. Similarly, more sophisticated \gls{MU} modelling methods may allow capturing important detail about the system's non-linearity. 

Like NeuroMotion \cite{ma_neuromotion_2024}, we also assumed rigid tendons in our \gls{MSk} model. While for low-force behaviours, such as gesturing, this is a common simplification \cite{hicks_is_2015}, it is worth considering incorporating it for more dynamic interactions or movements. As movement is performed, the length of the musculotendon units changes alongside the muscle morphology; this has an effect on the \gls{MUAP} templates measured at the electrodes. Although we used static templates, a natural extension would be dynamic \gls{MUAP} generation. This may be achieved either through a discretised lookup table or by integrating a regressor such as BioMime directly in the environment \cite{ma_conditional_2025}. 

\gls{MUAP} shapes are also influenced by physiological parameters, which we kept fixed throughout our experiments. Instead, these could be varied systematically alongside anthropometrics to create a distribution of potential virtual users. This type of domain randomisation is a promising technique that could provide a diversity necessary to train robust \glspl{HMI} that have the potential for zero- or few-shot transfer to real-life conditions \cite{akkaya_solving_2019}. 

Our synthetic \gls{EMG} surrogate signal was entirely dependent on the state of the \gls{MU} activity. In reality, many sources of noise and interference have significant influence over the signal, e.g., electrode noise, motion artefacts, electrocardiogram contamination and cross-talk \cite{de_luca_filtering_2010}. Our initial experiments with white Gaussian noise should be extended by including characterised noise models that allow training and evaluation with these detrimental signal elements.

\subsubsection{Limitations in the learning}
The gesturing task was a suitable first choice as a case study. However, our methods should be repeated with other motor control tasks. Reaching and manipulation would be reasonable next steps, to cover situations with distinct sub-motions and interactions with objects. On the other hand, simpler, wrist or thumb control tasks would be suitable to tackle intent estimation in more challenging signal conditions, such as low signal-to-noise ratio \gls{EMG}, or if crosstalk is considered with spatial \glspl{MUAP}. Future work should also investigate further options for physiologically inspired sensory inputs to the control policy, such as mechanoreceptors or Golgi-tendon organs. These may be key to induce realistic impedance control in the virtual user. Inputs to the \gls{EMG} decoder may be extended by other signal modalities commonly used in \gls{HMI} devices, including inertial measurement units that can be readily simulated \cite{hodossy_high-level_2025}.

\gls{PPO} is often considered a best-in-class \gls{RL} algorithm to tackle novel learning problems due to its robustness and flexibility \cite{schulman_proximal_2017}. However, its sample inefficiency makes it unsuitable to represent shorter-term motor learning and adaptation of feedforward control. Incorporating hierarchical control, supervised learning or recurrence mechanisms are good opportunities to tackle this \cite{park_generative_2022, kadiallah_generalization_2012, akkaya_solving_2019}. Lastly, we are currently missing the opportunity to leverage the differentiable physics available in MJX. Being able to differentiably relate actions to the changes in rewards can accelerate learning \cite{ren_diffmimic_2023}. Alternatively, this property could be exploited to integrate Jacobian-based control methods or allow fitting anthropometric or physiological parameters to experimental data.

We avoided the additional complication of simultaneous, multi-agent learning, by always updating one learning system at a time. Simultaneous learning is of great interest for future work, as co-adaptive \glspl{HMI} may be needed to maintain high performance over time without recalibration sessions.

\subsection{Applications}
High-throughput, closed-loop neuromechanical simulations may serve as much more than a theoretical exercise. As a development environment, they can partially alleviate the burden of neurophysiological experiments for developing and testing \gls{HMI} control algorithms. In combination with more accurate in silico \gls{MU} models, they could model both the biosignals and the motor behaviour (e.g., tremors) associated with neurological conditions, and help characterise key parameters or develop interventions such as exoskeletons or electrical stimulation strategies. Beyond virtual prototyping, through sufficiently diverse domain randomisation, cross-user generalisation may be possible in simulation. This would indicate a significant opportunity for sim2real transfer of user-agnostic \gls{HMI} controllers.

\section{Conclusion}

In this study we presented an open-source framework for simulating closed-loop neural control tasks in the hand. We leveraged parallelized reinforcement learning techniques to learn a control policy from the virtual user, trained an intent estimator offline to decode the synthetic \gls{EMG} signals. Lastly, we demonstrated that the human-device performance can be improved if the user can further adapt to the decoder during use. The developed tools are released as an open-source framework, laying the foundation for generating neural signals in realistic closed-loop scenarios in a tractable manner, which may be used to augment datasets to make them more robust or to cover pathological conditions.

Simulation provide a valuable platform to test our understanding, find answers and iterate on designs of the next generation of neural interfaces. If issues are identified and resolved within these cycles, evidence based choices can be made earlier without the need of a full experimental study. By no means does this eliminate the need for validation and quantification with full scale user studies. It does, however, allow for more efficient use of the limited resources available to conduct them.
\begin{ack}
This work was funded by the Imperial-Meta Wearable Neural Interfaces Research Centre. We would like to express our thanks to Claudia Sabatini and Dimitrios Chalatsis for the valuable discussions during this study.
\end{ack}
\bibliography{ref}
\appendix

\section{MU activity and EMG synthesis} \label{app:EMG}
Adopting a convolutive \gls{EMG} model, to synthesise plausible \gls{EMG} signals two components are necessary:
\begin{enumerate}
    \item \textbf{A process to determine \gls{MU} discharge times that correspond to a motor behaviour of interest.} In turn, this process involves making decisions regarding the properties of the common input and independent noise \cite{farina_accessing_2012}, and how these (primarily continuous) inputs are integrated to map to the sparse \gls{MU} firing activity. In-silico models traditionally focus on motor behaviours with neural drives that are easier to characterise, e.g., isometric contraction \cite{pascual-valdunciel_personalized_2025}. These conditions are well suited to tune and study properties of \gls{MU} models, however, most activities of daily living involve dynamic and non-stationary motor behaviours.
    \item \textbf{A model that associates a \gls{MUAP} to each \gls{MU}.} A \gls{MU} discharge evokes a wave of electric potential field that varies over space and time \cite{merletti_tutorial_2019}. Most commonly \glspl{MUAP} are described in discrete samples across these two variables, providing a time series specific for each \gls{EMG} channel and subject. One approach is to experimentally determine a collection of \glspl{MUAP} through decomposition techniques \cite{del_vecchio_tutorial_2020}, which may be reused inside simulations. Alternatively, finite element models or their approximations may be used to estimate these templates. \gls{MUAP} models may be conditioned on physiological, anatomical, and morphological parameters. As with the \gls{MU} firing, assumptions of isometric tasks are often made in these models \cite{maksymenko_myoelectric_2023}. This partially avoids the issue of the varying relative position of the electrode and the \gls{MU} \cite{guerra_wearable_2024}. Like before, this assumption constrains investigation to more static behaviours. Kinetic and kinematic factors may be included in the model to cover dynamic contractions \cite{ma_conditional_2025}.   
\end{enumerate}

To generate \gls{MU} firing, we use a leaky-integrate-and-fire model \cite{teeter_generalized_2018}. In our framework we allow selecting a subset of muscles of interest to simulate \glspl{MU} for them, and how many \glspl{MU} should contribute to the \gls{EMG} signals recorded at each electrode (we chose to simulate 220 significantly contributing \glspl{MU} in total).

In our current implementation, we use static \gls{MUAP} shapes; in practice these shapes warp as a function of kinematics and kinetics, a property that would be a potential improvement for future development. BioMime was used to generate the \gls{MUAP} in a neutral pose \cite{ma_conditional_2025}. We implement support for two types of \glspl{MUAP} stored at 1 kHz sampling: Spatial \glspl{MUAP} that account for cross-talk at nearby electrodes suitable to represent \gls{sEMG}, and a no-cross-talk condition which approximates highly-specific, intramuscular recording conditions. In the following experiments we will always use the no-cross-talk condition as that allows the application of straightforward \gls{EMG} decoding models, simplifying our proof-of-concept. A limitation of our current process is that despite synthesizing the \gls{EMG} at 1 kHz, the motor unit activity and firing is updated only at 125 Hz. However, this is still more than twice the non-ballistic firing rates of \glspl{MU} \cite{pucci_maximal_2006, dideriksen_neural_2020}.

To generate the input to the \glspl{MU}, we simply assume it is a randomly initialised linear scaling from the corresponding muscle excitation value and a parametrised independent input noise. While this assumption slightly reduces the electromechanical delay, it is an improvement over previous work that used the muscle activation for the same purpose \cite{ma_neuromotion_2024}, as muscle excitation precedes activation. When an \gls{MU}'s activity level exceeds its firing threshold, the discharge time is recorded and it is set to a ``hyperpolarised'' state. The parameters used for the \gls{MU} modelling are configurable, and are available in the supplementary materials.

We implemented storing the firing history in a sparse representation, useful for regenerating the entire \gls{EMG} history for offline training of decoders in a memory-efficient way. For online, closed-loop control, we also exposed the option of storing a rolling window of \gls{EMG} signals, composed of the summed contributions from the \glspl{MU}.

\end{document}